\documentclass[prd,nofootinbib,amsfonts,notitlepage, 10pt]{revtex4-2}

\usepackage[utf8]{inputenc}
\usepackage[T1]{fontenc}
\usepackage{hyperref}
\usepackage{graphicx,bm,amsmath,color}
\usepackage[dvipsnames]{xcolor}
\usepackage{bbold}
\usepackage{wasysym}
\usepackage{verbatim}
\usepackage{amssymb, tensor}

\hypersetup{
colorlinks=true,         
linkcolor=Orange,          
citecolor=OliveGreen,           
urlcolor=Fuchsia   
}

\usepackage{multirow}
\usepackage{tikz}
\usepackage{graphicx}
\usepackage{float}
\usepackage{mathtools}
\usepackage{comment}

\DeclareMathOperator{\sech}{sech}

\DeclareMathOperator{\arctanh}{arctanh}
\makeatletter

\newcommand{\be}{\begin{equation}}
\newcommand{\ee}{\end{equation}}
\newcommand{\beq}{\begin{eqnarray}}
\newcommand{\eeq}{\end{eqnarray}}
\newcommand{\ba}{\begin{align}}
\newcommand{\ea}{\end{align}}

\newcommand{\mathi}{{\rm i}}

		\newcommand{\pderuno}[2]{\frac{\partial  #1}{\partial #2}}

  \usepackage{geometry}
\begin{document}

\title{Doubly special relativity as a non-local quantum field theory}
\author{J.J. Relancio}
\affiliation{\small Departamento de Matemáticas y Computación, Universidad de Burgos, Plaza Misael Bañuelos, 09001, Burgos, Spain\\Centro de Astropartículas y F\'{\i}sica de Altas Energ\'{\i}as (CAPA),
Universidad de Zaragoza, C. de Pedro Cerbuna, 12, 50009 Zaragoza, Spain}
\email{jjrelancio@ubu.es}
\author{L. Santamaría-Sanz}
\affiliation{\small Departamento de Física, Universidad de Burgos, Plaza Misael Bañuelos, 09001 Burgos, Spain}
\email{lssanz@ubu.es}

\begin{abstract}
In this work, we {explore the quantum theories} of the free massive scalar, the massive fermionic, and the electromagnetic fields, in a doubly special relativity scenario. This construction is based on a geometrical interpretation of the kinematics of {this} kind of theory. In order to describe the modified actions, we find that a higher (indeed infinite) derivative field theory is needed, from which the deformed kinematics can be read. From our construction we are able to restrict the possible models of doubly special relativity to {those} that preserve linear Lorentz invariance. We quantize the theories and also obtain a deformed version of the Maxwell equations. We analyze the electromagnetic vector potential for both an electric pointlike source and magnetic dipole. We observe that the electric and magnetic fields do not diverge at the origin for some models described with an {anti-de Sitter} {momentum} space but do for the de Sitter one in both problems. 
\end{abstract}

\maketitle

\section{Introduction}
A quantum gravity (QG) theory, merging quantum field theory (QFT) and general relativity (GR), has been sought for the last few decades. While there is not currently such a framework describing consistently both theoretical and phenomenological quantum effects of gravity, it is believed that the notion of spacetime should change at small scales/high energies. In these cases, the classical notion of spacetime should be replaced by a quantum one, with new consequences in the aforementioned regime. For example, in loop quantum gravity, the structure of spacetime takes the form of a spin foam \cite{Sahlmann:2010zf,Dupuis:2012yw}, while in causal set theory and string theory, a non-locality arises~\cite{Wallden:2013kka,Wallden:2010sh,Henson:2006kf,Mukhi:2011zz,Aharony:1999ks,Dienes:1996du}.

The search for such a QG theory is highly motivated by the fact that GR is perturbatively non-renormalizable~\cite{tHooft:1974toh,Goroff:1985sz,vandeVen:1991gw}. Since loop quantum gravity, causal set theory and string theory lack of well-defined and testable predictions, different ``top-down'' approaches to a fundamental QG theory have been considered. One possibility that has been widely studied in the literature is to add new terms proportional to the Ricci scalar in the Einstein-Hilbert action. Nevertheless, these so-called $f(R)$ theories~\cite{Sotiriou:2008rp} are not renormalizable. A different possibility is to consider terms proportional to the squared Ricci and Riemann tensors~\cite{Salvio:2018crh}. While these latter theories are perturbatively renormalizable~\cite{Stelle:1976gc}, they present an additional massive spin-2 ghost degree of freedom, causing the {Hamiltonian to be unbounded from below, which leads to vacuum instabilities of the physical system and to the loss of unitarity of the QFT}.  It is possible to solve this problem if, instead of considering some terms proportional to the Ricci scalar and Ricci and Riemann tensors, infinite derivatives of them are taken {into account~\cite{Biswas:2011ar}. Thus, non-local QFTs arise (note that non-local QFTs have already appeared before in other contexts~\cite{Efimov:1967pjn,Krasnikov:1987yj,Tomboulis:1997gg}). The considered functions of the derivatives depend only on the d'Alembertian, so covariance is preserved. Some of these approaches are known as infinite derivative gravity (IDG) theories~\cite{Modesto:2017sdr}. It is worth mentioning that, in these theories, the gravitational potential takes a finite value at the origin~\cite{ Frolov:2015bta,Boos:2018bxf,Buoninfante:2018stt,Buoninfante:2018xif,Buoninfante:2018rlq, Giacchini:2018wlf}, contrary to what happens in GR. Moreover, some recent work in IDG even suggests that the introduction 
of infinite derivative theories prevents black hole horizons~\cite{Koshelev:2017bxd,Buoninfante:2019swn}.} Starting from this scenario, it is possible to construct a QFT with infinite derivatives~\cite{Biswas:2011ar,Maggiore:2016gpx,Buoninfante:2018mre,Boos:2019fbu}. However, there is an arbitrariness in the choice of the functions of the derivatives. From string theory arguments, this function is an exponential~\cite{Mandelstam:1973jk,Kaku:1974fc,Green:1984fu,Witten:1985cc,Eliezer:1989cr,Biswas:2014tua},
from causal set theory it takes a non-analytical form~\cite{Aslanbeigi:2014zva,Belenchia:2014fda}, {and for other choices of IDG models, it is a polynomial in the ultraviolet limit (see \cite{Calcagni:2023goc} and references therein).}

A completely different approach was regarded in doubly/deformed special relativity (DSR)~\cite{Amelino-Camelia2002b,AmelinoCamelia:2001vy,AmelinoCamelia:2008qg}, considered as a low-energy limit of a QG which aims to predict phenomenological effects that can be detected nowadays. In this theory, the kinematics of special relativity (SR) is deformed by adding a high-energy scale. This {usually} modifies the dispersion relation,  as well as the conservation of energy and momentum, which is normally called composition law of momenta. While the former do not have to be modified, the latter must be, being this the main ingredient of this kind of theory. The {principal} feature of DSR theories is that a relativity principle is present, so there are some deformed Lorentz transformations making compatible the dispersion relation and the composition law. There are two {primary} kinematical models considered in the literature, $\kappa$-Poincaré~\cite{Lukierski:1991pn} and {Snyder kinematics~\cite{Battisti:2010sr}. However}, instead of considering that Poincaré invariance is deformed, a different possibility is a Lorentz invariance violation (LIV)~\cite{Colladay:1998fq,Kostelecky:2008ts}, where there is not a relativity principle, meaning that the laws of physics depends on the observer. In this scenario, there is a modification of the dispersion relation but not of the momentum conservation laws. As discussed in~\cite{Addazi:2021xuf,AlvesBatista:2023wqm}, the phenomenologies of both theories are quite different. 

Since DSR started to be considered at the beginning of this century, there are some topics that are not fully developed. In particular, there are only a few proposals to consider the electromagnetic (EM) interaction in this scheme~\cite{Harikumar:2011um,Ivetic:2019mkx}, compared with the more vast literature in the LIV scenario~\cite{Lammerzahl:2005jw,Bluhm:2008yt,Xiao:2010yx,Casana:2011du,Chang:2012ks,Pramanik:2012fj,Borges:2016uwl}. 
Moreover, while there is {an} effective field theory in LIV, known as standard model extension~\cite{Colladay:1998fq}, the complete formulation of a QFT in DSR is still missing, although there are several works in the literature trying to develop it~\cite{Dimitrijevic:2003wv,Freidel:2007hk,Meljanac:2010ps,Girelli:2010wi, Meljanac:2010et,Mercati:2010qhd,Meljanac:2011cs,
Meljanac:2017grw,Meljanac:2017ikx,Mercati:2018hlc,Franchino-Vinas:2018jcs,Franchino-Vinas:2019nqy,Arzano:2020jro,Meljanac:2021qgq,Franchino-Vinas:2021bcl}. In another vein, more efforts in DSR were focused on understanding the relationship between this kind of theory and a curved momentum space\footnote{The idea of a curved momentum space  was first considered by Born~\cite{Born:1938} in the 30's as a way to avoid the ultraviolet divergences in QFT.  Due to the success of renormalization, this idea was forgotten until some years ago, when it was considered as a possible path to quantum gravity.}~\cite{Girelli:2006fw,Amelino-Camelia:2014rga,Letizia:2016lew,Barcaroli:2015xda,Barcaroli:2016yrl,Barcaroli:2017gvg,Relancio:2022mia} (see~\cite{Kostelecky:2011qz,Barcelo:2001cp,Weinfurtner:2006wt,Torri:2021hpj} for the LIV geometrical description).  In the previous literature on the topic {it is not clear how to} relate in a simple way the composition of momenta with such momentum dependent metrics. A different point of view was developed in~\cite{Carmona:2019fwf}, finding that both the Lorentz transformations and the composition law must be isometries of the momentum metric, in order to have a relativity principle. {Since four   conservation laws for the momenta as well as Lorentz invariance} are required to describe a relativistic kinematics, a maximally symmetric momentum space is needed. From this construction, both $\kappa$-Poincaré  and Snyder kinematics can be easily obtained. Based on this last geometrical setup, in~\cite{Franchino-Vinas:2022fkh} an approach to a QFT in DSR, fully described in momentum space, was proposed. It is important to mention that the results obtained for the Dirac equation in that paper coincides with those of~\cite{Nowicki:1992if}, derived from an algebraic point of view.

The aim of this work is to translate and extend the results obtained in~\cite{Franchino-Vinas:2022fkh} for position space. Interestingly, we find an infinite derivative (and therefore non-local) field theory. In doing so, we are able to restrict the possible models of DSR. In particular, we find that the proposed model restricts the possible kinematics to those with a linear Lorentz invariance, i.e. Snyder models, with the Casimir being a function of the squared momentum.  While there are several (indeed infinite) choices of metrics in momentum space satisfying this property, we will focus on a very particular one obtained in~\cite{Relancio:2020rys,Chirco:2022jvx} from geometrical arguments.  {This does not imply, however, a total restriction on kinematics different from those of Snyder's models, which could be allowed in different proposals not considered here. Notice that for $\kappa$-Poincare kinematics in DSR, there exist bases for which linear Lorentz invariance is preserved at the level of single particles, while in other bases it is not. However, for a multi-particle system linear Lorentz invariance it is never preserved for these kinematics, whatever the basis one works on (see for instance \cite{Kowalski-Glikman:2002iba}). Regarding Snyder kinematics, all the considered bases possess linear Lorentz invariance for both single- and multi--particle systems (see \cite{Battisti:2010sr}). So, preservation of linear Lorentz invariance is not a restriction on DSR. As we will see in Sec.~\ref{sec:position}, we are imposing from the beginning that our model preserves linear Lorentz invariance at the single-particle level (where the Minkowski metric is present). Moreover,  a compatibility condition of the Klein--Gordon equation will impose that linear Lorentz invariance must be preserved for multi-particle systems. This is also interesting because, in this way, the one-particle results would be compatible with the algebraic scheme (see the discussion of \cite{Franchino-Vinas:2022fkh}), and because this could allow one to extend this work to a QFT considering multiple particle interactions too. Hence, our particular model is only compatible with Snyder kinematics. However, this does not mean that different models could be constructed in a compatible way with $\kappa$-Poincaré.  }

Moreover, {we also consider the EM Lagrangian in our construction of DSR QFT with infinite derivatives.\footnote{See~\cite{Accioly:2004sn} for a modification of the electric potential by considering a higher (finite) derivative EM Lagrangian.} From here}, we are able to describe the electric and magnetic potentials of a point charge and a magnetic dipole, respectively. We find that both potentials are finite at the origin (which is the case of the aforementioned IDG theories for the gravitational potential) only for the anti-de Sitter (AdS) model, seeming to privilege it with respect to the de Sitter space (dS)\footnote{{Note that whenever we refer to dS or AdS models we are dealing with curved momentum spaces.}}. This coincides with the results of~\cite{Matschull:1997du,Freidel:2003sp}, where the symmetries of a curved AdS momentum space, including a deformed composition of momenta, were obtained from a QG in $2+1$ dimensions coupled to point-like matter~\cite{Matschull:1997du,Freidel:2003sp}. 

The structure of the paper is as follows. {In Sec.~\ref{sec:revision} we review the main results of~\cite{Franchino-Vinas:2022fkh}}, building up the base of our description of QFT in DSR in position space,  {which will be carried out in Sec.}~\ref{sec:position}. We also show how our scheme can accommodate the field theories considered from string and causal set {scenarios}. In Secs.~\ref{sec:kg}-\ref{sec:dirac} we study the Klein--Gordon and Dirac equations, respectively, and their conserved energy-momentum tensors. From the former equation, we find that in order the dispersion relation to hold, the only kind{s} of kinematics allowed in this scheme are Snyder models. In Sec.~\ref{sec:em} we describe the EM {interactions, writing} the modified Maxwell equations,  and obtaining the electric and magnetic vector potentials of a point charge and a magnetic dipole, respectively.  Finally, we end with our conclusions and future prospects in Sec.~\ref{sec:conclusions}.

\section{Revisiting DSR QFT in momentum space}
\label{sec:revision}
Since~\cite{Franchino-Vinas:2022fkh} is inspired by previous works about geometrical interpretations of DSR, we start by reviewing the necessary {foundations} of the geometry of curved momentum spaces and their connection to relativistic deformed kinematics.   In~\cite{Carmona:2019fwf} it was shown that all the ingredients of a relativistic deformed kinematics can be obtained from a maximally symmetric momentum space. These core concepts are the deformed dispersion relation and the deformed composition law of momenta. The former establishes a relationship between the energy, momentum, and mass of a particle, and it can be identified  with the squared distance in momentum space. The latter describes the total momentum of a system of particles (which is {no longer} the sum of momenta but a non-linear function of them), and it can be obtained from translations in a curved momentum space.
Another crucial fact is that some Lorentz transformations (isometries of the maximally symmetric momentum space) make compatible the two previous ingredients, imposing then a relativity principle. These transformations, together with the deformed composition law,  establish a clear difference between DSR models and LIV theories.     

In~\cite{Relancio:2020zok} it was shown that the following relation between the metric and the Casimir\footnote{A Casimir operator is an operator which is not identical to the unit operator and which commutes with all group elements \cite{ Jonesbook}.
} (regarding it as the squared distance in momentum space) holds: 
  \begin{equation}
\frac{1}{4} \frac{\partial C(p)}{\partial p_\mu}g_{\mu \nu }(p)\frac{\partial C(p)}{\partial p_\nu}=C(p)\,,
\label{eq:casimir_metric}
  \end{equation}
  where one can define~\cite{Franchino-Vinas:2022fkh}:
  \begin{equation}
f^\mu  (p):=\frac{1}{2} \frac{\partial C(p)}{\partial p_\mu}\,.\label{eq:f_definition}
\end{equation}
For SR, the previous equation reduces to the well-known relation $C(p)=p_{\mu} \eta^{\mu \nu } p_\nu=m^2\,,$
where here and in the following we consider $\eta=\mathrm{diag}(1,-1,-1,-1)$.   In this way, the dispersion relation in the theory can be written as $C(p)-m^2=0$.

The Klein-Gordon and Dirac actions in the momentum space in DSR are given by:
\begin{align}\label{eq:DSR_actions_p}
S_{\rm KG}\,:&=\,\int {\rm d}^4p\, \sqrt{-g(p)} \, \tilde\phi^*(p) \left(  C(p)  -m^2  \right)   \tilde\phi(p)
\,,\\
    S_{\rm Dirac}:&=\!\int \!\!  {\rm d}^4p\, \sqrt{-g(p)}  \tilde \psi(-p)\left(  \gamma^\mu \eta_{\mu\rho} e^\rho{}_\nu({p}) f^\nu({p})-m\right) \tilde\psi(p)\,,
\end{align}
with $\tilde\phi(p)$ and $\tilde\psi(p)$ being the Fourier transforms of the scalar {and} fermionic fields, $\gamma^\mu$ the usual Dirac matrices in SR, and $ e^\mu{}_\nu(p)$ the tetrad in momentum space, which is related to the metric by means of 
$g_{\mu\nu}(p) =e^\rho{}_\mu (p)\eta_{\rho \sigma}e^\sigma{}_\nu (p)\,.$ {Given a composition law $\oplus$, the tetrad can be determined in the following way~\cite{Carmona:2019fwf}:
\be
{e}^\mu{}_\nu (p):= \left. \frac{\partial \left(p \oplus q\right)_\nu}{\partial q_\mu}\right|_{q\to 0}\,.
\label{eq:tetrad2}
\ee  
In this way, different kinematics obtained from the same metric lead to different tetrads and hence, to different field theories~\cite{Franchino-Vinas:2022fkh}.}

\section{Toward a DSR QFT in position space}
\label{sec:position}
{In this section, we discuss how to consider the  QFT in position space corresponding to the momentum space version in the previous section. We also discuss the possible momentum metrics allowed in our scheme, and how this scheme can accommodate the non-local field theories obtained from string and causal set theories.}

\subsection{Construction of our approach}\label{subsec3a}
In order to generalize QFT so that DSR deformed symmetries are taken into account, our proposed construction is to replace the usual derivatives in QFT with a function of them that takes into account the curvature in momentum space, so the deformed dispersion relation is satisfied. Thus, we consider the following action for scalar fields (see the fermionic version in {Sec. }\ref{sec:dirac}):
\begin{eqnarray}
&&\!\!\!\!\!\!S=\!\!\int {\rm d}^4 x \frac{1}{2} \left\lbrace   -\ell^\mu(-i\partial_x)\,  \phi(x) \eta_{\mu \nu} \ell^\nu(-i\partial_x)\,  \phi(x)  -m^2 \phi^2(x) \right\rbrace\,,
\label{eq:KG_action}
   \end{eqnarray}
where  
\begin{equation}
\ell^\mu (-i\partial_x)= {e^\mu}_\nu (p) f^\nu (p)\left.\right|_{p \to -i\partial_x}\,.
\end{equation}
This modification leads to a non-local QFT, since an infinite number of derivatives act on the fields. Moreover, linear Lorentz invariance is preserved. {This} restricts the possible bases of kinematics to those whose dispersion relation is a quadratic expression on the momentum, so that\footnote{In the following, for alleviating the notation, when functions are expressed in terms of $p^2$ we mean $p^2/\Lambda^2$, for dimensional reasons.}
\begin{equation}
f^\mu  (p)=\frac{1}{2} \frac{\partial p^2}{\partial p_\mu}\frac{\partial C(p)}{\partial p^2}=p^\mu \frac{\partial C(p)}{\partial p^2} \coloneqq p^\mu h(p^2)\,.
   \label{eq:f_cov}
\end{equation}
Notice that having a relativity principle implies that Lorentz transformations and the composition law must be isometries of the momentum metric. This happens whenever the metric is of the form
\begin{equation}
g_{\mu\nu}(p)=\eta_{\mu\nu} \, \Theta_1 (p^2)+\frac{p_\mu p_\nu}{\Lambda^2}\,  \Theta_2(p^2)\,,
\label{eq:lorentz_metric}
\end{equation}
where $\Lambda$ is the {high-energy} scale. This kind of metric was also found to be privileged from a geometrical point of view when a curvature in both momentum space and spacetime is considered~\cite{Relancio:2020rys,Pfeifer:2021tas}. There are different (indeed, infinite) choices of momentum metrics that possess linear Lorentz transformations as isometries. In the next subsection we discuss some particular options.

{But before going on, we would like to discuss the differences and similarities between  our approach and that of~\cite{Dias:2016lkg,deBrito:2016fqe}. In these works, starting from a particular deformed uncertainty relation, the authors define a deformed momentum operator. This operator, which is a nonlinear function of the usual momentum, is used to replace the derivatives in the associated QFT. Then, a higher-derivative QFT is constructed in this  way. In our approach, the starting point is a metric in momentum space and from it we derive a  $\ell^\mu$ operator, which replaces the usual momentum operator. While we can redefine this operator to obtain some desired deformed uncertainty relations, it is more difficult to obtain the specific choice of metric which leads to the uncertainty relations of~\cite{Dias:2016lkg,deBrito:2016fqe}. This fact could constitute an obstacle whenever one wants to consider a deformed kinematics in the DSR context. Nevertheless, the connection between the two approaches opens a new line of research which we hope to deepen in the future. }

\subsection{Possible choice of metrics}
\label{sec:pcm}
As mentioned in the Introduction, a non-local action in QFT was obtained from string theory. In particular, the interaction terms {involving} infinite derivatives are of the form~\cite{Mandelstam:1973jk,Kaku:1974fc,Green:1984fu,Witten:1985cc,Eliezer:1989cr,Biswas:2014tua}
\begin{equation}
    V(\phi)\sim \left(e^{-\alpha \Box }\right)^3\,,
\end{equation}
where $\Box=\partial^\mu\partial_\mu$ as usual, and $\alpha$ is a constant which depends on whether the string is open or closed, and on the Regge slope. Inspired by this theory,  several works of IDG and non-local QFT arose ~\cite{Frolov:2015bta,Buoninfante:2018stt,Buoninfante:2018xif,Buoninfante:2018rlq,Buoninfante:2018mre}{. They consider} the action in~\eqref{eq:KG_action} together with the choice
\begin{equation}
    C(-i\partial_x)=-\Box e^{\Box/\Lambda^2 }\,, \label{CasimirIDG}
\end{equation}
or, equivalently, 
\begin{equation}
    C(p)=p^2 e^{-p^2/\Lambda^2 }\,.
\end{equation}
Let us see how this proposal for the Casimir operator can be embedded in our formalism. It can be viewed as a particular example of our construction based on a curved momentum space. In particular, from the condition in~\eqref{eq:casimir_metric} one finds the following relationship between the functions defining the metric in~\eqref{eq:lorentz_metric}:
\begin{equation}
    \Theta_2 (p^2)= \frac {\Lambda^2}{p^2} \left(\frac{e^{p^2/\Lambda^2}\Lambda^4}{(p^2-\Lambda^2)^2}-  \Theta_1 (p^2)\right)\,.
    \label{relation_thetas}
\end{equation}
Imposing that the momentum metric must  correspond to a maximally symmetric space (see Appendix~\ref{sec:appendix}), one finds that 
\begin{eqnarray}
 &&   \Theta^\text{dS}_1 (p^2)= -\frac{p^2}{\Lambda^2} \sech^2 \left(\frac{\sqrt{p^2}}{\Lambda}e^{-p^2/\Lambda^2}\right)\,,  \qquad \textrm{and}
    \qquad 
    \Theta_1^\text{AdS} (p^2)= \frac{p^2}{\Lambda^2} \sec^2 \left(\frac{\sqrt{p^2}}{\Lambda}e^{-p^2/\Lambda^2}\right)\,, 
\label{eq:thetaads}
\end{eqnarray}
for dS and AdS cases, respectively. As we see, the Minkowski metric is not recovered when taking the limit $\Lambda \to \infty$. This means that the composition of momenta cannot be written as a power series expansion when considering it as an isometry of the metric~\cite{Carmona:2019fwf} i.e., when writing
\begin{equation}
g_{\mu\nu}\left(p\oplus q\right) \,=\,\frac{\partial \left(p\oplus q\right)_\mu}{\partial q_\rho} g_{\rho\sigma}(q)\frac{\partial \left(p\oplus q\right)_\nu}{\partial q_\sigma}\,,
\label{eq:composition_isometry}
\end{equation} 
making it difficult to search for phenomenological implications. 

A different choice for the Casimir operator could be the one obtained in causal set theory~\cite{Aslanbeigi:2014zva,Belenchia:2014fda}, whose expression at lower orders in $\Lambda$ is given by
\begin{equation}
    C(-i\partial_x)=-\Box -\frac{3}{2\pi \sqrt{6}}\frac{\Box^2}{\Lambda^2}\left(3 \gamma -2+\ln \left(\frac{3 \Box^2}{2 \pi \Lambda^4}\right)\right)+\cdots\,,
\end{equation}
with $\gamma$ being {the Euler-Mascheroni constant}. In this case one could obtain also the function $\Theta_2$, which defines the metric but only order by order, since this expression is not analytic. But since it is possible to obtain the associated metric and impose that it is maximally symmetric even if it is order by order, in principle it can be a viable choice for our construction. 

In the following, we will focus on a very particular case, in which the maximally symmetric momentum metric is conformally Minkowski~\cite{Relancio:2020rys,Chirco:2022jvx}:
\begin{equation}
g_{\mu\nu}(p)=\eta_{\mu \nu}\left(1\pm\frac{p^2}{4\Lambda^2}\right)^2\,,
\label{eq:conformal_metric}
\end{equation}
where the  plus sign stands for AdS and the minus for dS. This metric was obtained from geometrical considerations inside the DSR scheme when a curvature in spacetime was also considered, being hugely special due to its conformal form. Due to the arbitrariness present in our construction, we will develop the QFT with this metric (although we compare some results with the  Casimir of IDG in~\eqref{CasimirIDG}). It could be the case that, by extending  our proposal to curved spacetimes in future works, the previous metric appears to be privileged, as it is in the geometrical setup. 

In the next few sections, we will consider \eqref{eq:conformal_metric}, together with the QFT scheme described in {Sec.} \ref{subsec3a}, in order to generalize the Klein--Gordon, Dirac, and electromagnetic Lagrangians.  

\section{Klein-Gordon equation}
\label{sec:kg}
In this section we start by considering the Klein-Gordon equation in position space. While it could seem very naive and trivial, we find several interesting results. From the action in position space, we conclude that the only way in which the Klein-Gordon equation can be obtained in our scheme is by selecting Snyder models over $\kappa$-Poincaré kinematics. Moreover, we study the conserved quantities associated with this equation, finding a deformation of the usual result of QFT. Notice that in this section we have already promoted the classical problem to a quantum one. Consequently, now the momentum $p$ must be treated as an operator $p\to -i\partial_x$. We will abuse notation by using the same letters for the quantum variables and the classical ones, because they can be distinguished by the context.

\subsection{Action in position space}
Consider for simplicity real scalar fields. Let us compute the small variations of the functional~\eqref{eq:KG_action} with respect to the scalar field $\phi$:
\begin{align}\label{Svar}
& \delta S_\phi\!=\!\int \! {\rm d}^4 x \,\, \frac{1}{2}  \left\{ -\ell^\mu(-i\partial_x)\,  \delta \phi(x) \, \eta_{\mu\nu}\, \ell^\nu(-i\partial_x) \, \phi(x)
 -  \ell^\mu(-i\partial_x) \phi (x)\, \eta_{\mu\nu}\,  \ell^\nu(-i\partial_x) \, \delta \phi(x) -2m^2 \phi(x) \, \delta \phi(x)\right\}.
 \end{align}
Notice that by integrating by parts as follows
\begin{eqnarray}\label{intbyparts}
&& \int\! {\rm d}^4 x \left[\ell(-i\partial_x)\phi(x)\right] \psi(x)
=\!\int \!{\rm d}^4 x  \left[ \sum_{n=0}^{\infty} \frac{\ell^{(n)}(0)}{n!} \left(-i \partial_x\right)^n \phi(x)\right] \psi(x)
=\sum_{n=0}^{\infty} \frac{\ell^{(n)}(0)}{n!} \int {\rm d}^4 x \left[ (-i \partial_x)^n \phi(x)\right] \psi(x)
 \nonumber\\ 
&&=\!\sum_{n=0}^{\infty} \frac{\ell^{(n)}(0)}{n!} \int\! {\rm d}^4 x  \, \phi(x) \left[ (i \partial_x)^n \psi(x)\right]=\int {\rm d}^4 x \, \phi(x) \left[\ell\left( i \partial_x\right) \psi(x)\right] ,
\end{eqnarray}
we can rewrite \eqref{Svar} as
\begin{align}
\delta S\,=&\,\int {\rm d}^4 x \frac{1}{2}  \left\{ -  \delta \phi(x) \,  \ell^\mu(i\partial_x)\ell_\mu(-i\partial_x) \, \phi(x)
 -  \ell_\mu(i\partial_x) \ell^\mu(-i\partial_x) \phi (x) \, \delta \phi(x) -2m^2 \phi(x) \, \delta \phi(x)\right\}.
\end{align}
Applying the variational principle yields the equation of motion:
\begin{eqnarray}
\left(    \ell^\mu(i\partial_x)\ell_\mu(-i\partial_x) 
 +m^2 \right)\phi(x)=0.
\end{eqnarray}

The only way in which the dispersion relation $C(p)-m^2=0$ holds is by imposing $\ell^\mu(-i\partial_x)=-\ell^\mu(i\partial_x)$. From Eq.~\eqref{eq:f_cov}, it is easy to see that $f^\mu(-p)=-f^\mu(p)$, so  $e^\mu{}_\nu(-p)=e^\mu{}_\nu(p)$ must be fulfilled. {This means that the composition law leading to the tetrad following~\eqref{eq:tetrad2} cannot correspond to   $\kappa$-Poincaré},\footnote{{Note that the composition law in $\kappa$-Poincaré kinematics always depends on a fixed vector $n^\mu$, see~\cite{Kowalski-Glikman:2002iba}. Then, the corresponding tetrad computed through Eq.~\eqref{eq:tetrad2} will lead to $e^\mu{}_\nu(n,p)\neq e^\mu{}_\nu(n,-p)$ (see a similar discussion carried out in~\cite{Franchino-Vinas:2022fkh}).}} leaving only space for Snyder models. Thus we can replace $\ell^\mu(-i\partial_x) \to-i \partial^\mu \,\Omega(-\Box):\equiv -i\tilde{\partial^\mu}$ in the previous equations, with  $\Omega$ being a function depending on the momentum space tetrad, the function of the d'Alembertian $h(-\Box)$, and the deformed dispersion relation. 

Taking into account the previous notation for $\ell^\mu$, the  Klein--Gordon action for real fields is 
\begin{equation}
S=\int {\rm d}^4 x\,\,  \frac{1}{2}\left\lbrace  \tilde \partial^\mu \phi(x) \tilde \partial_\mu \phi(x)  -m^2 \phi^2(x) \right\rbrace\,.
\label{eq:KG_action3}
   \end{equation}
Notice that the Lagrangian density of  a free complex scalar field theory  will be \begin{equation}\mathcal{L}=\tilde \partial^\mu  \phi^*(x) \tilde \partial_\mu \phi(x)  -m^2 \phi^*(x) \phi(x) \,. 
\end{equation}
From now on we will consider the real scalar fields case for simplicity, but the generalization to complex ones could be performed straightforwardly.

 Notice that from the metric~\eqref{eq:conformal_metric}, one finds  the following tetrad:
\begin{equation}
{e^\mu}_\nu(p)=\delta^\mu_\nu\left(1\pm\frac{p^2}{4\Lambda^2}\right)\,,
\label{eq:conformal_tetrad}
\end{equation}
where again the plus sign stands for AdS and the minus for dS, and the dispersion relations become~\cite{Relancio:2020rys}
\begin{equation}
C_\text{dS}(p)\,=\,4 \Lambda^2 \arctanh^2\left(\frac{\sqrt{p^2}}{2 \Lambda}\right)=m^2\,, \qquad C_\text{AdS}(p)\,=\,4 \Lambda^2 \arctan^2\left(\frac{\sqrt{p^2}}{2 \Lambda}\right)=m^2\,,
\label{eq:casimir_ds}
\end{equation}
with $p^2=p_0^2-\vec{p}^{\, 2}$. Since $h(p^2)$ is the derivative of the Casimir with respect to the four-momentum squared, we can compute the exact form of the $\Omega$ function in both models: 
\begin{eqnarray}
\ell^\mu(-i\partial_\mu)&=& \left.\delta^\mu_\nu\left(1\pm\frac{p^2}{4\Lambda^2}\right)  p^\nu \, h(p^2)\right|_{p_\mu \to -i\partial_\mu}=\left. \left(1\pm\frac{p^2}{4\Lambda^2}\right)  p^\mu \, \pderuno{C(p)}{p^2}\right|_{p_\mu \to -i\partial_\mu}=\left. -i\partial^\mu \, \Omega(p^2)\right|_{p_\mu \to -i\partial_\mu}\, ,\nonumber\\
\Omega_\text{dS}(-\Box)&=&\left(1+\frac{\Box}{4\Lambda^2}\right)\frac{8 \Lambda^3}{\sqrt{-\Box} \,\, (4 \Lambda^2+\Box)}\arctanh\left(\frac{\sqrt{-\Box}}{2 \Lambda}\right)\,, \\
\Omega_\text{AdS}(-\Box)&=&\left(1-\frac{\Box}{4\Lambda^2}\right)\frac{8 \Lambda^3}{\sqrt{-\Box} \,\, (4 \Lambda^2-\Box)}\arctan\left(\frac{\sqrt{-\Box}}{2 \Lambda}\right)\,.
\label{h_ads-ds}
\end{eqnarray}

\subsection{Energy-momentum tensor and conserved currents}
The energy-momentum tensor can be computed as~\cite{Birrell:1982ix}
\begin{equation}
T_{\mu\nu}=\frac{2}{\sqrt{-\mathtt{g}(x)}}\frac{\delta \mathcal{L}}{\delta \mathtt{g}^{\mu\nu}(x)}\,.
    \label{eq:energy_momentum_tensor}
\end{equation}
Note that here we use $\mathtt{g}$ for describing the metric of spacetime (depending on spacetime coordinates), and this should not be confused with the momentum dependent metric $g$. When taking the limit for flat spacetime, {i.e. $\mathtt{g}_{\mu\nu} (x) \to \eta_{\mu\nu}$}, we are able to find the energy-momentum tensor for our theory. Indeed, we obtain the usual conservation law of this tensor:
\begin{equation}
\partial^\mu T_{\mu\nu}=0\,.
\label{eq:conservation_em_tensor}
\end{equation}

When considering the Lagrangian density of the action~\eqref{eq:KG_action}, one finds (see Appendix~\ref{sec:appendixB})
\begin{equation}
    T_{\mu\nu}= \tilde{\partial}_\mu \phi  \tilde{\partial}_\nu \phi -  \frac{1}{2}\eta_{\mu\nu}\left[ \tilde{\partial}^\rho \phi(x)  \tilde{\partial}_\rho \phi(x)  -m^2  \phi^2(x) \right]\,.
    \label{eq:energy_momentum_kg}
\end{equation}
From this expression, it is easy to check that the conservation law~\eqref{eq:conservation_em_tensor} is satisfied. 

A simple way to define the Fourier transform of the field in terms of annihilation and creation operators, as in usual QFT \cite{Itzykson:1980rh} and in high-derivative theories \cite{Buoninfante:2018mre}{,} is 
\begin{equation}
   \phi(x)= \int  \frac{{\rm d}^3 p }{(2\pi)^{3}\sqrt{2 p_0}}\,  \,\left(a_\mathbf{p} e^{-\mathi x^\lambda p_\lambda}+a^\dagger_\mathbf{p} e^{\mathi x^\lambda p_\lambda} \right)\,.
   \label{eq:KG_field2}
\end{equation}

Using the definition of Eq.~\eqref{eq:KG_field2} in Eq.~\eqref{eq:energy_momentum_kg}, and after a little algebra, one can obtain the conserved currents, which are (see Appendix~\ref{sec:appendixB})
\begin{equation}
    T_{0\mu}=\frac{s(m^2)}{2}\int \frac{\text{d}^3 p}{(2\pi)^3} p_\mu\left(a_p^\dagger a_p+a_p a^\dagger_p\right)\,.
    \label{eq:conserved_em_kg}
\end{equation}
Notice that $s(m^2)= \Omega(C^{-1}(m^2))$ is a function of the mass, which depends on the model considered and its Casimir. We find that the energy and the momenta are conserved. {Moreover, } instead of using~\eqref{eq:KG_field2}, one can introduce a normalization factor 
\begin{equation}
   \phi(x)= \int  \frac{{\rm d}^3 p }{(2\pi)^{3}\sqrt{2 s(m^2) p_0}}\,  \,\left(a_\mathbf{p} e^{-\mathi x^\lambda p_\lambda}+a^\dagger_\mathbf{p} e^{\mathi x^\lambda p_\lambda} \right)\,,
   \label{eq:KG_field3}
\end{equation}
so the conserved currents are the same of those of standard QFT
\begin{equation}
    T_{0\mu}=\frac{1}{2}\int \frac{\text{d}^3 p}{(2\pi)^3} p_\mu\left(a_p^\dagger a_p+a_p a^\dagger_p\right)\,.
\end{equation}

\section{Dirac equation}
\label{sec:dirac}
In this section we discuss the Dirac equation in spacetime, computing the corresponding {energy-momentum} tensor. 

\subsection{Derivation from an action}
 Consider the following  action for fermionic fields:
\begin{align}
 S&=\int {\rm d}^4 x\,   \bar\psi(x)\left(-\gamma^\mu\, \eta_{\mu\nu} \, \ell^\nu (-i\partial_x)-m\right)\psi(x)= \int {\rm d}^4 x\,   \bar\psi(x)\left(i \gamma^\mu\, \tilde\partial_\mu -m\right)\psi(x)\,.
 \label{eq:Dirac_action}
\end{align}
Computing variations, and using~\eqref{intbyparts} together with $\bar\psi(x)=\psi^{\dagger}(x)\gamma^0$, one obtains 
\begin{eqnarray}
\delta S_{\psi}&=&  \int {\rm d}^4 x\, \left\lbrace \delta\bar\psi(x)\left(-\gamma^\mu\,  \ell_\mu (-i\partial_x)-m\right)\psi(x)+\psi^{\dagger}(x)\gamma^0\left(-\gamma^\mu\,  \ell_\mu (-i\partial_x)-m\right)\delta\psi(x)\right\rbrace 
\nonumber\\
&=&\int {\rm d}^4 x\, \left\lbrace \delta\bar\psi(x)\left(-\gamma^\mu\,  \ell_\mu (-i\partial_x)-m\right)\psi(x)+\ell_\mu(i\partial_x)  \psi^{\dagger}(x)\gamma^0\left(-\gamma^\mu  -m\right)\delta\psi(x)\right\rbrace
\nonumber\\
&=&\int {\rm d}^4 x\, \left\lbrace \delta\bar\psi(x)\left( -\gamma^\mu\,  \ell_\mu (-i\partial_x)-m\right)\psi(x)+\ell_\mu(i\partial_x)  \psi^{\dagger}(x)\left( -\gamma^0 \gamma^\mu \gamma^0   -m\right)\gamma^0 \delta\psi(x)\right\rbrace
\nonumber\\
&=&\int {\rm d}^4 x\, \left\lbrace \delta\bar\psi(x)\left( -\gamma^\mu\,  \ell_\mu (-i\partial_x)-m\right)\psi(x)+\ell_\mu(i\partial_x)  \psi^{\dagger}(x)\left( -\gamma^\mu{}^\dagger    -m\right)\gamma^0 \delta\psi(x)\right\rbrace
\nonumber\\
&=&\int {\rm d}^4 x\, \left\lbrace \delta\bar\psi(x)\left(i \gamma^\mu  \tilde \partial_\mu -m\right)\psi(x)+\left[\left(i \gamma^\mu  \tilde \partial_\mu-m\right)\psi(x)\right]^{\dagger}\gamma^0 \delta\psi(x)\right\rbrace\,,
\end{eqnarray}
where we  have the same equation  twice. If one considers only the fermionic case, no restriction on the possible kinematics is obtained from the equations arising from the variational principle. However, since we are interested in a full QFT, both for fermions and bosons, we will choose the kinematics bases respecting Lorentz invariance (in particular, the Snyder models), as in the scalar case, because this is the most restrictive case.

\subsection{Conserved currents}
We can also construct the energy-momentum tensor in this case. As shown in~\cite{Birrell:1982ix}, Eq.~\eqref{eq:energy_momentum_tensor} can be rewritten in terms of the tetrad, resulting in
\begin{equation}
T_{\mu\nu}=\frac{\mathtt{e}_{\alpha\mu}}{\text{det}[\mathtt{e}(x)]}\frac{\delta \mathcal{L}}{\delta {\mathtt{e}^{\nu}}_\alpha(x)}\,.
\label{eq:energy_momentum_tensor_tetrad}
\end{equation}
As before, the tetrad of the spacetime $\mathtt{e}$ (depending on the spacetime coordinates) should not be confused with the momentum space tetrad  $e$ considered before. By taking the flat spacetime limit of this expression, and plugging the Lagrangian involved in~\eqref{eq:Dirac_action} into~\eqref{eq:energy_momentum_tensor_tetrad}, one obtains (see Appendix~\ref{sec:appendixB})
\begin{equation}
    T_{\mu\nu}=i \bar\psi(x)\gamma_\mu\tilde \partial_\nu \psi(x)-\eta_{\mu \nu}\bar\psi(x)\left(i\gamma^\rho \tilde \partial_\rho -m\right) \psi(x)\,.
    \label{eq:energy_momentum_dirac}
\end{equation}
Now, if we quantize the fermionic field as usual
\begin{align}
   \psi(x)&= \int  \frac{{\rm d}^3 p }{(2\pi)^{3}\sqrt{2 p_0}}\, \sum_{s=1,2} \,\left(a_{\mathbf{p},s}u^{(s)}(\mathbf{p}) e^{-\mathi x^\lambda p_\lambda}+b^\dagger_{\mathbf{p},s}v^{(s)}(\mathbf{p}) e^{\mathi x^\lambda p_\lambda} \right),\\
      \psi^\dagger(x)&= \int  \frac{{\rm d}^3 q }{(2\pi)^{3}\sqrt{2 q_0}}\, \sum_{r=1,2} \,\left(b_{\mathbf{q},r}v^{(r)\dagger}(\mathbf{q}) e^{-\mathi x^\lambda q_\lambda}+a^\dagger_{\mathbf{q},r}u^{(r)\dagger}(\mathbf{q}) e^{\mathi x^\lambda q_\lambda} \right)\,,
   \label{eq:diracq}
\end{align}
one finds the conserved quantities (see Appendix~\ref{sec:appendixB})
\begin{eqnarray}
    P_\mu=s(m^2) \int \frac{{\rm d}^3 p}{(2\pi)^3} \, p_\mu  \sum_{s=1,2} \left(b^\dagger_{\mathbf{p},s} b_{\mathbf{p},s} +a^\dagger_{\mathbf{p},s} a_{\mathbf{p},s}\right)\, .
\end{eqnarray}
where again $s(m^2)= \Omega(C^{-1}(m^2)).$

\section{Electromagnetic Lagrangian}
\label{sec:em}
Following the formalisms described in the Introduction, the deformed Maxwell tensor in our DSR-QFT theory must be
\begin{equation}
\tilde{F}_{\mu\nu}=i \ell_\mu (-i\partial_x) A_\nu -i \ell_\nu(-i\partial_x) A_\mu= \tilde \partial_{\mu}  A_{\nu}-\tilde \partial_{\nu} A_{\mu} = \Omega(-\Box) F_{\mu\nu}\,,
   \label{eq:Maxell_tensor}
\end{equation}
with $F_{\mu\nu}$ being the usual electromagnetic tensor. We see that gauge invariance holds for the transformation given by $A^\prime_\mu=A_\mu+\tilde \partial_\mu  \theta\,. $
Then, the usual Maxwell Lagrangian is generalized to 
\begin{align}
 S&=-\int  {\rm d}^4 x \frac{1}{4}\tilde{F}_{\mu\nu}\tilde{F}^{\mu\nu}\,.
 \label{eq:action_EM}
\end{align}
The associated Lagrangian can be written as 
\begin{equation}
-\frac{1}{4}\tilde{F}_{\mu\nu}\tilde{F}^{\mu\nu}=-\frac{1}{2}\left(\tilde \partial_{\mu}   A_{\nu}\tilde \partial^{\mu}  A^{\nu}-\tilde \partial_{\nu}  A_{\mu}\tilde \partial^{\mu}  A^{\nu}\right)\,.
   \label{eq:FF}
\end{equation}
When making small variations of the action with respect to $A_\mu$ one finds 
\begin{equation}
i \ell^\mu(-i\partial_x)\tilde{F}_{\mu\nu}= \tilde \partial^\mu  \tilde{F}_{\mu\nu}=0\,,
\label{eq:derivative_F}
\end{equation}
and, when replacing \eqref{eq:Maxell_tensor} in the above expression together with the Lorentz gauge $\partial^\mu  A_\mu=0\,,$
every component of the electromagnetic vector satisfies the Klein-Gordon equation, viz.
\begin{equation}
C(-\Box)A_\nu=-\Box \, \Omega^2(-\Box)  A_\nu= 0\,.
\end{equation}
Moreover, from Eq.~\eqref{eq:derivative_F}, it is easy to see  that the electric and magnetic fields defined as usual, i.e.
\begin{equation}
E_i=\partial_0 A_i-\partial_i A_0\,, \qquad  B_i=\epsilon_{i j k } \partial_j A_k\,,
\end{equation}
with $\epsilon_{ijk}$ being the Levi-Civita tensor, leads to a deformation of the Maxwell equations. For that aim, when considering in Eq.~\eqref{eq:derivative_F} $\nu=0$ one finds
\begin{equation}
 \Omega^2(-\Box) \partial^\mu \left(\partial_\mu A_0-\partial_0 A_\mu\right)= \Omega^2(-\Box)\left( \partial^i \partial_i A_0-\partial^i \partial_0 A_i\right) = -\Omega^2(-\Box)\partial^i E_i=0\,,
\end{equation}
and for $\nu=i$
\begin{equation}
\Omega^2(-\Box) \partial^\mu \left(\partial_\mu A_i-\partial_i A_\mu\right)= \Omega^2(-\Box)\left( \partial^0 \left(\partial_0 A_i-\partial_i A_0\right)  -\partial^j \left(\partial_j A_i - \partial_i A_j\right) \right) = \Omega^2(-\Box)\left(\partial^0 E_i+\epsilon_{ijk}\partial_j B_k\right)=0\,.
\end{equation}
From the previous expressions one can write the deformed Maxwell equations in a vectorial form
\begin{equation}
  \Omega^2(-\Box) \nabla\cdot\vec E=0\,,\qquad  \Omega^2(-\Box) \vec \nabla\times \vec B= \Omega^2(-\Box) 
 \partial_0 \vec E\,.
 \label{eq:maxwell1}
\end{equation}
On the other hand, as in standard EM theory, one can obtain the other two deformed Maxwell equations as before 
\begin{equation}
 \Omega^2(-\Box) \vec \nabla\cdot\vec B=0\,,\qquad \Omega^2(-\Box) \vec \nabla\times  \vec E=-\Omega^2(-\Box)\partial_0 \vec B\,,
  \label{eq:maxwell2}
\end{equation}
from the dual tensor 
\begin{equation}
\tensor[^*]{\tilde{F}}{_\mu_\nu}:=\frac{1}{2}\epsilon_{\mu\nu\rho\sigma}\tilde{F}_{\rho \sigma}\,,
\end{equation}
where $\epsilon_{\mu\nu\rho\sigma}$ is the rank-4 Levi-Civita symbol. By applying the operator $\Omega^2(-\Box)$ on the identity
\begin{equation}
\vec \nabla\times  (\vec \nabla\times \vec E )=\vec \nabla\cdot (\vec \nabla\cdot \vec E) -\vec \nabla^2 \vec E\,,
\label{eq:identity}
\end{equation}
and taking into account Eqs.~\eqref{eq:maxwell1} and \eqref{eq:maxwell2}, one finds
\begin{eqnarray}
\Omega^2(-\Box) \vec \nabla\times  (\vec \nabla\times \vec E )&=& -\Omega^2(-\Box)\vec \nabla^2 \vec E\,,\nonumber\\
-\Omega^2(-\Box) \vec \nabla\times\partial_0  \vec B &=& -\Omega^2(-\Box)\vec \nabla^2 \vec E\,\nonumber\\
-\Omega^2(-\Box) 
 \partial_0\partial_0 \vec E &=& -\Omega^2(-\Box)\vec \nabla^2 \vec E\,.
\end{eqnarray}
When a similar computation is carried out for the magnetic field, the Klein--Gordon equation for both fields are obtained
\begin{equation}
C(-\Box) \vec E=0\,, \qquad C(-\Box)\vec B =0\,.
\label{eq:ceyb}
\end{equation}
The electromagnetic energy-momentum tensor obtained from ~\eqref{eq:energy_momentum_tensor} and~\eqref{eq:action_EM} can be written as (see Appendix~\ref{sec:appendixB})
\begin{equation}
    T_{\mu\nu}= \frac{1}{4}\eta_{\mu\nu} \tilde{F}_{\rho \sigma}\tilde{F}^{\rho \sigma}+\tilde{F}_{\mu \rho} {\tilde F^\rho}_{\,\,\,\,\nu}\,.
\end{equation}
The energy and the Pointing vector associated with this energy-momentum tensor are the same as those of standard QFT, once~\eqref{eq:ceyb} is taken into account
\begin{equation}
    T_{00}= \frac{1}{2}(\vec E^2+\vec B^2)\,,\qquad  T_{0i}= (\vec E \times \vec B)_i\,.
\end{equation}

Now, we can add a minimal coupling to matter to the EM Lagrangian, finding 
\begin{align}
 S_{EM}&=-\int  {\rm d}^4 x \left(\frac{1}{4}\tilde{F}_{\mu\nu}\tilde{F}^{\mu\nu}+j^\mu A_\mu\right)\,.
 \label{eq:action_EM2}
\end{align}
When varying the action with respect to the electromagnetic vector potential one finds 
\begin{equation}\label{eomcharge}
-C(-\Box)A^\mu=j^\mu\,.
\end{equation}
Moreover, it is easy to find the deformed Maxwell equations with an external source, which are~\eqref{eq:maxwell2} together with
\begin{equation}
\Omega^2(-\Box) \nabla\cdot\vec E=j_0\,,\qquad  \Omega^2(-\Box) \vec \nabla\times \vec B= \Omega^2(-\Box)   \partial_0 \vec E + \vec j\,.
 \label{eq:maxwell3}
\end{equation}
Using now Eqs.~\eqref{eq:maxwell2}, \eqref{eq:identity} and \eqref{eomcharge}, the Klein--Gordon equation for the fields are obtained
\begin{equation}
C(-\Box) \vec E=\nabla j_0+\partial_0 \vec j\,, \qquad C(-\Box)\vec B =-\vec\nabla\times \vec j\,.
\end{equation}

 \subsection{Electric field of a point charge at the origin}
\begin{figure}[H]
    \centering
    \includegraphics[width=0.5\textwidth]{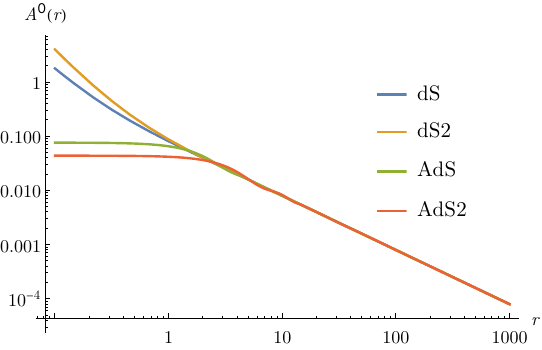}
    \caption{\small Electric scalar potential $A^0(r)$ as a function of the radius $r$ for different de Sitter and anti-de Sitter models and $q=\Lambda=1$.}
    \label{fig:A0rto0}
\end{figure}

We now consider the electric field of a point-like source. In this case, for the potential  $j^0(x)=q \, \delta^3(\vec{r}), \, j^i=0$, the e.o.m. turn out to be exactly analytically solvable. The 0-component of the vector potential is the electric scalar potential whose Fourier transform (FT) can be written as
\begin{eqnarray}\label{A0def}
A^0(\vec{r}) = \frac{1}{(2\pi)^3} \int {\rm d}^3k\,  \tilde{A^0}(k) \, e^{i \vec{k}\cdot \vec{r}}\, .
\end{eqnarray}
Consequently, the equation of motion for this scalar field is 
\begin{eqnarray}
    -C(-\Box)A^0 &=&q \, \delta^3(\vec{r}),\nonumber\\
     -\frac{1}{(2\pi)^3} \int {\rm d}^3k\,  \tilde{A^0}(k) \, C(-\vec{k}^2)\, e^{i \vec{k}\cdot \vec{r}}&=&\frac{q}{(2\pi)^3} \int {\rm d}^3k\,  e^{i \vec{k}\cdot \vec{r}}\,.
\end{eqnarray}
After comparing the two sides of the equation above, it can be seen that $\tilde{A^0}(k) = -q/C(-\vec{k}^2)$. Plugging this FT component into the definition of the field \eqref{A0def} and integrating in spherical coordinates yields
\begin{eqnarray}\label{neweq65}
    A^0(r) &=& -\frac{q}{(2\pi)^3} \int  \! \frac{{\rm d}^3k\, e^{i \vec{k}\cdot \vec{r}} }{C(-\vec{k}^2)} = - \frac{q}{(2\pi)^3} \int_0^{2\pi} \!\!\! \int_0^\pi \!\! \int_0^\infty \frac{k^2 \sin^2\theta \, {\rm d}k\, {\rm d}\theta {\rm d}\varphi }{C(-\vec{k}^2)} \, e^{i k r \cos \theta} =- \frac{q}{(2\pi)^2} \int_0^\infty \! \frac{{\rm d}k\, k^2}{C(-\vec{k}^2)} \int_{-1}^{1} \! {\rm d}u\,  e^{i k r u} \nonumber\\
    &=& -\frac{q}{2\pi^2} \int_0^\infty {\rm d}k\, \frac{ k^2}{C(-\vec{k}^2)}  \frac{\sin(kr)}{kr}\, .
\end{eqnarray}
In \cite{ourletter} it is shown that the electric scalar potential is constant for the AdS scenario and divergent for the dS one when $r\to 0$. This behavior can also be seen in FIG. \ref{fig:A0rto0}, where we have compared the $A^0(r)$ obtained from~\eqref{eq:casimir_ds} with the value arising when considering the Casimir operators of the classical basis of the {Snyder model in the Maggiore representation~\cite{Franchino-Vinas:2022fkh}}:
\begin{eqnarray}\label{Casimirs1}
    C_\text{dS2}(-\vec{p}^2)= - \Lambda^2 \text{arcsinh}^2 \left(\frac{\sqrt{\vec{p}^2}}{\Lambda}\right), \qquad  C_\text{AdS2}(-\vec{p}^2)= - \Lambda^2 \text{arcsin}^2 \left(\frac{\sqrt{\vec{p}^2}}{\Lambda}\right)\,.
\end{eqnarray}In addition,  whenever one chooses $C=e^{-\nabla^2/ \Lambda^2} \, \nabla^2$, the electric potential is given by
\begin{eqnarray}\label{Casimirs2}
    A^0_\text{IDG}(r)=  \frac{q}{2\pi^2} \int_0^\infty dk\,  \frac{\sin(kr)}{kr\,  e^{k^2/\Lambda^2}}=\frac{q}{4\pi}Erf\left(\frac{r}{2}\right),
\end{eqnarray}
with $Erf$ being the error function. The behavior of this last electric scalar potential at $r \to 0$ is analogous to that of the AdS scenario, {as shown in \cite{ourletter}}. 

Concerning the electric field, it can be obtained by means of $\vec{E}=-\vec{\nabla} A^0(r)$, i.e.
\begin{equation}
\vec{E}=\frac{q}{2\pi^2} \int_0^\infty {\rm d}k\, \frac{ k^2}{C(-\vec{k}^2)}\frac{k r \cos (k r)-\sin (k r)}{k r^2}\, \Vec{u}_r\, .
\label{eq:electric_field}
\end{equation}
An important conclusion that can be easily checked is that the electric field does not diverge at $r \to 0$ for the AdS and IDG models but does for the dS ones. {As discussed in Sec.~\ref{sec:pcm}, the operator of IDG is not compatible with a composition law such that, when making a series power expansion in $\Lambda$, the zeroth-order term corresponds to the usual sum of momenta. This means that it should not be considered within our model since we want to recover the special relativistic case when $\Lambda \to \infty$. However, this does not mean that there are no different constructions for which IDG is valid. Moreover, from a mathematical point of view, in order to compute the electromagnetic potentials in Eq. \eqref{neweq65} one only needs a Casimir, independently of the form of the metric (and its corresponding composition of momenta) and its physical implications. Therefore, we have also computed the IDG case (which is well known in the literature and shows a finite behavior of potentials at the origin of coordinates) in order to compare this result with the proposed Casimirs allowed in our construction.  }

 \subsection{Magnetic field of a magnetic moment at the origin}

\begin{figure}[h]
    \centering
    \includegraphics[width=0.45\textwidth]{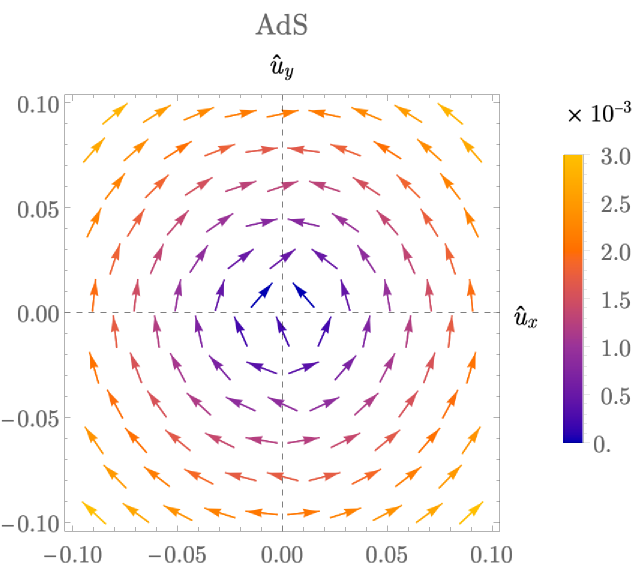}\qquad \includegraphics[width=0.45\textwidth]{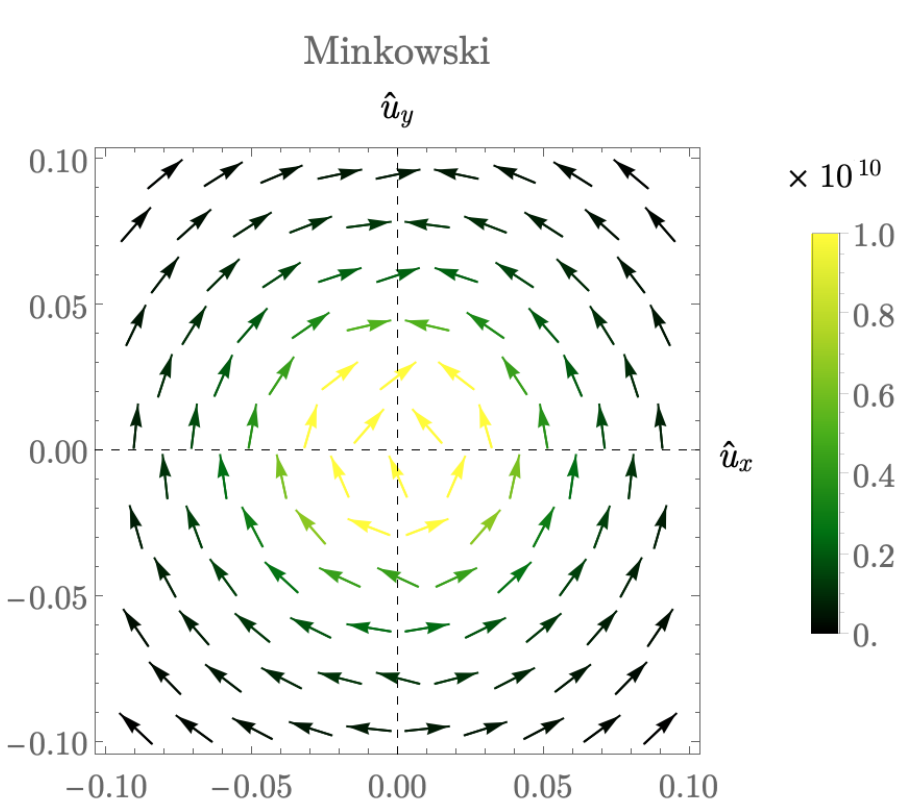}
    \caption{Magnetic vector potential $\Vec{A}(r)$ for $\vec{m}=(0,0,1)$ and $\Lambda=1$ in the AdS (left) and Minkowski (right) models for small values of $r$.}
    \label{fig:Amag}
\end{figure}

\begin{figure}[h]
    \centering
    \includegraphics[width=0.45\textwidth]{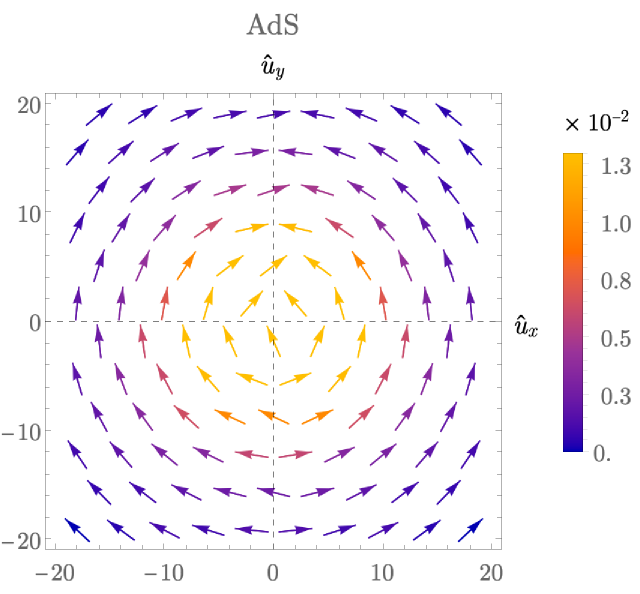}\qquad \includegraphics[width=0.45\textwidth]{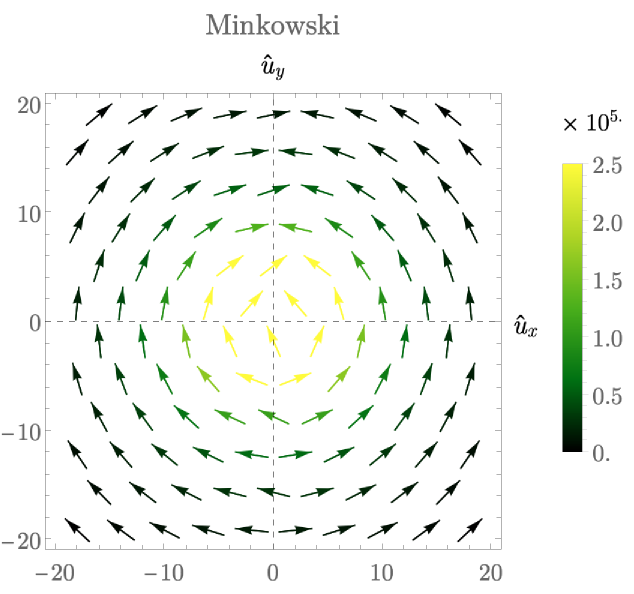}
    \caption{Magnetic vector potential $\Vec{A}(r)$ for $\vec{m}=(0,0,1)$ and $\Lambda=1$ in the AdS (left) and Minkowski (right) models for higher values of $r$.}
    \label{fig:Amag1}
\end{figure}

 If one considers a static magnetic moment $\Vec{m}$ localized at the origin~\cite{Jackson:1998nia}, i.e. $j^0(x)=0, \, \vec{j}=-\Vec{m} \times \Vec{\nabla}\delta^3(\Vec{r})$, the $i$-components of the vector potential constitute the magnetic vector potential, whose Fourier transform can be written as
\begin{eqnarray}\label{Adef}
\Vec{A}(\vec{r}) = \frac{1}{(2\pi)^3} \int {\rm d}^3k\,  \tilde{\Vec{A}}(k) \, e^{i \vec{k}\cdot \vec{r}}\,.
\end{eqnarray}
 Now, Eq. \eqref{eomcharge} turns into 
\begin{equation}
-C(\nabla^2)\, \Vec{A}(\vec{r})=-\Vec{m} \times \Vec{\nabla}\delta^3(\Vec{r}) = \Vec{\nabla} \times \Vec{m}\, \delta^3(\Vec{r})\, .
\end{equation}
From here, the Fourier transform of the magnetic vector potential can be obtained
\begin{eqnarray}
-C(\nabla^2)\, \Vec{A}(\bar{r})\, &=& \Vec{\nabla} \times \Vec{m}\, \delta^3(\Vec{r})\,,\nonumber\\
-\frac{1}{(2\pi)^3} \int {\rm d}^3k\,  \tilde{\Vec{A}}(k) \, C(-\vec{k}^2)\, e^{i \vec{k}\cdot \vec{r}}&=& \Vec{\nabla} \times \frac{\Vec{m}}{(2\pi)^3}\,\int {\rm d}^3k\,  e^{i \vec{k}\cdot \vec{r}}= \frac{1}{(2\pi)^3}\int {\rm d}^3k\,\,   \Vec{\nabla} \times \left( \Vec{m} \,e^{i \vec{k}\cdot \vec{r}}\right)\,.
\end{eqnarray}
Consequently,
\begin{equation}
\tilde{\Vec{A}}(k) = -\frac{ e^{i \vec{k}\cdot \vec{r}}}{C(-\vec{k}^2)}\, \Vec{\nabla} \times \left( \Vec{m} \, e^{i \vec{k}\cdot \vec{r}} \right)\,. 
\end{equation}
The magnetic vector potential is thus
\begin{eqnarray}
    \Vec{A}(r) &=& -\frac{1}{(2\pi)^3} \int   \frac{{\rm d}^3k}{C(-\vec{k}^2)}\, \Vec{\nabla} \times \left( \Vec{m} \, e^{i \vec{k}\cdot \vec{r}} \right)= \frac{\Vec{m}}{(2\pi)^3}  \times  \int \!\!   \frac{{\rm d}^3k}{C(-\vec{k}^2)}\, \Vec{\nabla} e^{i \vec{k}\cdot \vec{r}}= \frac{\Vec{m}}{(2\pi)^2}  \times \vec{\nabla} \int_0^\pi \!\!\int_0^\infty \frac{k^2 \sin \theta  {\rm d}\theta {\rm d}k}{C(-\vec{k}^2)} e^{i k r \cos \theta}\nonumber\\
   &=&\frac{\Vec{m}}{2\pi^2}  \times  \vec{\nabla} \int_0^\infty \frac{k^2  {\rm d}k}{C(-\vec{k}^2)} \frac{2\sin(kr)}{kr} = \Vec{m}  \times \frac{\Vec{r}}{2\pi^2r^3}\int_0^\infty \frac{k \, {\rm d}k}{C(-\vec{k}^2)} \left( kr \cos(kr)-\sin(kr)\right):= \Vec{m} \times [\Vec{r}\, f(r)]\,.
\end{eqnarray}
It is straightforward to see from here that the magnetic vector potential for AdS is zero in the limit $r \to 0$ (see FIG. \ref{fig:Amag}). This feature also applies for the IDG model considered in \eqref{Casimirs2}. The only difference between the two models is the numerical value obtained for $\Vec{A}(r)$, although the order of magnitude remains the same. On the contrary, the magnetic vector potential diverges at the origin for the undeformed model (i.e. Minkowski spacetime) as well as for the dS case. It is easy to check that for higher values of the distance to the origin, all the models (dS, AdS and IDG) recover the characteristic behavior of the Minkowski space (see FIG. \ref{fig:Amag1}), in which the vector potential becomes zero at a large distance from the origin.

In order to compute the magnetic field $\Vec{B}= \Vec{\nabla}\times  \Vec{A}(r)$, one basically needs to calculate the rotational of the magnetic vector potential. In this way:
\begin{eqnarray}
\Vec{\nabla}\times \Vec{A}(r) &=& \Vec{\nabla}\times \left[\Vec{m} \times (\Vec{r} f(r))\right] = \Vec{m}\Vec{\nabla}(\Vec{r} f(r))-(\Vec{m}\Vec{\nabla})(\Vec{r} f(r))=  \Vec{m}\frac{1}{r^2} \partial_r(r^3 f(r))-m_r \partial_r (rf(r)) \Vec{u}_r- m_\theta f(r)\Vec{u}_\theta\nonumber\\
&&-m_\varphi f(r)\Vec{u}_\varphi = \Vec{m}  \left( 2 f(r)+ r \partial_r f(r)\right) -\frac{\Vec{m} \Vec{r}}{r} \partial_rf(r) \Vec{r}.
\end{eqnarray}
Similarly to what happens when evaluating the electric field of a point charge when approaching the origin, the magnetic field goes to zero for the AdS and IDG models but diverges for the dS one.

   \section{Conclusions and outlook}
\label{sec:conclusions}
In this work we make a new proposal for considering QFT in DSR theories. This construction is built upon a geometrical setup used in a previous work regarding QFT in momentum space. Here, we discuss how this formalism can be translated to position space. This leads to a non-local QFT, since infinite derivatives of the fields appear. Furthermore, we show that any non-local QFT can in principle be embedded in our scheme,  showing explicitly how to do so for the particular example considered in string theory. 

In our construction of QFT, we find that not every basis of DSR kinematics is allowed, but only those with a dispersion relation which is a function of the momentum squared. Moreover, when deriving the Klein--Gordon equation, we also find that only Snyder kinematics can be considered, impeding us from working in $\kappa$-Poincaré models. We have also generalized the Dirac equation and the electromagnetic Lagrangian within our scheme.

Due to the arbitrariness present in our theory (there are infinite Snyder models), we limit ourselves to a particular basis previously considered in the literature as privileged by a geometrical argument. The corresponding metric is conformally Minkowski, and the only possible arbitrariness comes down to whether we consider a de Sitter or anti-de Sitter momentum space. For both models, we are able to compute the electric and magnetic fields produced by a charged point particle and a magnetic dipole, respectively.  We find that only for AdS the potentials do not diverge at zero, privileging this metric over dS, since non-local theories are considered as models that regularize this kind of divergence. Our finding is in agreement with previous results in the literature claiming that a quantum gravity theory in 2+1 dimensions has the symmetries of an  AdS momentum space.

Since the electromagnetic and gravitational potentials are computed in a similar way, the gravitational field of a point-like mass will present the same regular behavior at the origin. Therefore, if one were to calculate the gravitational potential one would also find that it does not diverge at the origin.  This means that our construction of a QFT from a geometrical setup is also a good candidate for a non-local theory of gravity and should also be considered as a different possible and interesting model. The extension of our proposal for considering gravity will be carried out in a future work. Moreover, as commented before, it could be the case that this extension explicitly shows that our choice of metric is privileged over other metrics, as is the case in previous works in the literature regarding DSR in curved spacetimes. 


Finally, we would like to point out some caveats about phenomenology. DSR was developed as a low-energy limit of a QG theory with phenomenological implications. These possible effects were looked for from astrophysical implications, since very high energies should be reached in order for these quantum gravity corrections to appear. The perspective proposed in this work opens a new phenomenology based on tabletop experiments (see also~\cite{Mercati:2010au} for other proposals of experiments using slow atoms). For example, in~\cite{Belenchia:2015ake,Belenchia:2016zaa} it was shown that the deformed dispersion relation considered in causal set theory leads to a modified Schrödinger evolution in the non-relativistic limit. This  leads to  a very characteristic effect that cannot be generated by the environment, so in principle it can be tested in the laboratory. Therefore, this work opens a new branch of phenomenology of DSR at low energies that could also be tested, and used to restrict the high-energy scale depicting the model, in a parallel way to astrophysical experiments.

\section*{Acknowledgments}
This work has been supported by the grant PID2023-148373NB-I00 funded by MCIN /AEI /10.13039/501100011033 / FEDER, UE, and by the Q-CAYLE Project funded by the Regional Government of Castilla y Le\'on (Junta de Castilla y Le\'on) and by the Ministry of Science and Innovation (MCIN) through the European Union funds NextGenerationEU (PRTR C17.I1). The authors thank J. L. Cortés and S. Liberati for fruitful discussions. JJR is funded by the Agencia Estatal de Investigación (Spain) through grant PID2019-106802GB- I00/AEI/10.13039/501100011033. The authors thank J. L. Cortés and S. Liberati for fruitful discussions.

\appendix
\section{Scalar of curvature of the momentum space}
\label{sec:appendix}

In this appendix we show how to obtain the functions defining the metric for the Casimir of IDG.  We start by the definition of the curvature tensor in momentum space~\cite{miron2001geometry}
\begin{equation}
{S_{\sigma}}^{\mu\nu\rho}(p)\,=\, \frac{\partial C^{\mu\nu}_\sigma(p)}{\partial p_\rho}-\frac{\partial C^{\mu\rho}_\sigma(p)}{\partial p_\nu}+C_\sigma^{\lambda\nu}(p) \,C^{\mu\rho}_\lambda(p)-C_\sigma^{\lambda\rho}(p)\,C^{\mu\nu}_\lambda(p)\,,
\label{eq:Riemann_p}
\end{equation} 
 where 
 \begin{equation}
C_\rho^{\mu\nu}(p)\,=\,\frac{1}{2}g_{\rho\sigma}(p) \left(\frac{\partial g^{\sigma\nu}(p)}{\partial p_ \mu}+\frac{\partial g^{\sigma\mu}(p)}{\partial p_ \nu}-\frac{\partial g^{\mu \nu}(p)}{\partial p_ \sigma}\right)\,,
\label{eq:affine_connection_p}
\end{equation}
is the momentum affine connection. Since it it a maximally symmetric space, the following equation holds
\begin{equation}
S^{\tau \mu \nu \kappa}(p)\,=\,\pm \frac{1}{\Lambda^2}\left( g^{\tau \nu}(p)g^{\kappa \mu}(p)-g^{\tau \kappa}(p)g^{\mu \nu}(p)\right)\,,
\label{eq:momentum_gs}
\end{equation} 
where the positive sign is for dS and the negative one for AdS. While this expression was obtained for space-time metrics~\cite{Weinberg:1972kfs}, it is easy to check that also holds for momentum-space metrics. 

By substituting the relationship between the functions defining the metric obtained in Eq.~\eqref{relation_thetas}, and using Eq.~\eqref{eq:momentum_gs}, one finally obtains a set of differential equations of second order in $\Theta_1$. These can be easily solved leading to expressions in~\eqref{eq:thetaads}. 

\section{Energy momentum tensor and conserved currents}
\label{sec:appendixB}
\subsection{Scalar field theory}
The Lagrangian for scalar fields with a generic metric is given by
\begin{eqnarray}
    \mathcal{L}= \frac{\sqrt{-\mathtt{g}}}{2} \left(\tilde{\partial}_\mu \phi(x) \, \mathtt{g}^{\mu\nu}(x) \, \tilde{\partial}_\nu \phi(x)-m^2\phi^2(x)\right)\,.
\end{eqnarray}
Performing small variations of the action with respect to $\mathtt{g}$ one obtains
\begin{eqnarray}
\delta \mathcal{L}= -\frac{1}{2} \frac{\sqrt{-\mathtt{g}}}{2} \,  \mathtt{g}_{\mu\nu}(x)\delta\mathtt{g}^{\mu\nu}(x)\left(\tilde{\partial}_\rho \phi(x) \, \mathtt{g}^{\rho\sigma}(x) \, \tilde{\partial}_\sigma \phi(x)-m^2\phi^2(x)\right)+ \frac{\sqrt{-\mathtt{g}}}{2} \tilde{\partial}_\mu \phi(x)\, \delta\mathtt{g}^{\mu\nu}(x) \, \tilde{\partial}_\nu \phi(x)\, ,
\end{eqnarray}
where we have used~\cite{Birrell:1982ix}
\begin{equation}
    \delta \sqrt{-\mathtt{g}}=-\frac{1}{2}\sqrt{-\mathtt{g}}\,\mathtt{g}_{\mu\nu}\delta \mathtt{g}^{\mu\nu}\,.\label{vardetg}
\end{equation}
Consequently, the energy-momentum tensor can be written as
\begin{eqnarray}
    T_{\mu\nu}&=& \left.\frac{2}{\sqrt{-\mathtt{g}}} \, \frac{\delta\mathcal{L}}{\delta \mathtt{g}^{\mu\nu}(x)}\right|_{g_{\mu\nu} \to \eta_{\mu\nu}} =\left. \left[-\frac{1}{2}  \,  \mathtt{g}_{\mu\nu}(x)\left(\tilde{\partial}_\rho \phi(x) \, \mathtt{g}^{\rho\sigma}(x) \, \tilde{\partial}_\sigma \phi(x)-m^2\phi^2(x)\right)+ \tilde{\partial}_\mu \phi(x) \,  \tilde{\partial}_\nu \phi(x)\right]\right|_{g_{\mu\nu} \to \eta_{\mu\nu}} \nonumber\\
&=& \tilde{\partial}_\mu \phi(x) \,  \tilde{\partial}_\nu \phi(x)-\frac{1}{2}  \,  \eta_{\mu\nu}(x)\left(\tilde{\partial}^\sigma \phi(x) \, \tilde{\partial}_\sigma \phi(x)-m^2\phi^2(x)\right)\,.
\end{eqnarray}
From here, the conserved quantities can be computed. If one replaces 
\begin{eqnarray}
    \phi(x)= \int \frac{{\rm d}^3p}{(2\pi)^3 \sqrt{p_0}} \left(a_p e^{-\mathi x^\lambda p_\lambda} + a_p^\dagger e^{\mathi x^\lambda p_\lambda} \right)
\end{eqnarray}
in the expression for $ T_{0\mu}$, one obtains:
\begin{eqnarray}
  &&P_\mu=  \int {\rm d}^3 x\,  T_{0\mu} =- \int \frac{{\rm d}^3p}{(2\pi)^3 \sqrt{p_0}} \frac{{\rm d}^3q}{(2\pi)^3 \sqrt{q_0}} \, p_0 \, q_\mu \, \Omega(p^2)\, \Omega(q^2)\left(a_p a_q e^{-\mathi x^\lambda (q_\lambda+p_\lambda)} + a_p^\dagger a_q^\dagger e^{\mathi x^\lambda (q_\lambda+p_\lambda)} \right.\nonumber\\
&&\left.\hspace{30pt}  -a_p^\dagger a_q e^{-\mathi x^\lambda (q_\lambda-p_\lambda)} - a_p a_q^\dagger e^{-\mathi x^\lambda (-q_\lambda+p_\lambda)}  \right)= \int \frac{{\rm d}^3p}{2(2\pi)^3 }  \,   p_\mu \, \Omega^2(p^2)\left(a_p^\dagger a_q  + a_p a_q^\dagger \right)\, ,
\end{eqnarray}
where we have used the integral definition of $\delta^3(p-q)$, as well as the fact that the first two terms in the summation are zero because they result from the integration of an odd function $a_p a_{-p}$ (similarly $a^\dagger_p a^\dagger_{-p}$) on a symmetric interval. 

\subsection{Fermionic field theory}
The Lagrangian for Dirac fields with a generic metric is given by
\begin{eqnarray}
    \mathcal{L}= \sqrt{-\mathtt{g}} \, \bar\psi(x)\left(i \gamma^\alpha\, {\mathtt{e}^\nu}_\alpha (x)\tilde\partial_\nu -m\right)\psi(x)={\text{det}[\mathtt{e}(x)]} \, \bar\psi(x)\left(i \gamma^\alpha\, {\mathtt{e}^\nu}_\alpha (x)\tilde\partial_\nu -m\right)\psi(x)\,.
\end{eqnarray}
Performing small variations of the action with respect to $\mathtt{e}$ one obtains
\begin{eqnarray}
\delta \mathcal{L}&=& \delta{\text{det}[\mathtt{e}(x)]} \, \bar\psi(x)\left(i \gamma^\beta\, {\mathtt{e}^\sigma}_\beta (x)\tilde\partial_\sigma -m\right)\psi(x)+ {\text{det}[\mathtt{e}(x)]} \, \bar\psi(x)i \gamma^\alpha\, \delta{\mathtt{e}^\nu}_\alpha(x)\tilde\partial_\nu \psi(x)=\nonumber\\
&=&-\text{det}[\mathtt{e}(x)] \, {\mathtt{e}^\alpha}_\nu \, \delta {\mathtt{e}^{\nu}}_\alpha \, \bar\psi(x)\left(i \gamma^\beta\, {\mathtt{e}^\sigma}_\beta (x)\tilde\partial_\sigma -m\right)\psi(x)+ {\text{det}[\mathtt{e}(x)]} \, \bar\psi(x)i \gamma^\alpha\, \delta{\mathtt{e}^\nu}_\alpha(x)\tilde\partial_\nu \psi(x)\, .
\end{eqnarray}
Notice that we have used~\eqref{vardetg} together with $\mathtt{g}_{\mu\nu}= {\mathtt{e}^\alpha}_\mu \, \eta_{\alpha \beta} \, {\mathtt{e}^\beta}_\nu$ to compute the variation of the determinant of the tetrad in the first term. Now, the energy-momentum tensor can be written as
\begin{eqnarray}
    T_{\mu\nu}&=& \left.\frac{\mathtt{e}_{\mu\alpha}}{\text{det}[\mathtt{e}(x)]}\frac{\delta \mathcal{L}}{\delta {\mathtt{e}^{\nu}}_\alpha(x)}\right|_{{\mathtt{e}^{a}}_b \to {\delta^{a}}_b}=\left. -{\mathtt{e}^\rho}_{\alpha}\,  \eta_{\rho \mu}\, {\mathtt{e}^\alpha}_\nu \, \bar\psi(x)\left(i \gamma^\beta\, {\mathtt{e}^\sigma}_\beta (x)\tilde\partial_\sigma -m\right)\psi(x)+ i\mathtt{e}_{\mu\alpha} \, \bar\psi(x)\, \eta^{\alpha \beta}\gamma_\beta\, \tilde\partial_\nu \psi(x)\right|_{{\mathtt{e}^{a}}_b \to {\delta^{a}}_b}\nonumber\\
    &=&\left. -{\delta^\rho}_{\nu}\,  \eta_{\rho \mu}\, \bar\psi(x)\left(i \gamma^\beta\, {\mathtt{e}^\sigma}_\beta (x)\tilde\partial_\sigma -m\right)\psi(x)+ i{\mathtt{e}^\beta}_{\mu} \, \bar\psi(x)\, \gamma_\beta\, \tilde\partial_\nu \psi(x)\right|_{{\mathtt{e}^{a}}_b \to {\delta^{a}}_b}\nonumber\\&=& i \bar\psi(x)\gamma_\mu\tilde \partial_\nu \psi(x)-\eta_{\mu \nu}\bar\psi(x)\left(i\gamma^\alpha \tilde \partial_\alpha -m\right) \psi(x)\,,
\end{eqnarray}
because ${\mathtt{e}^\alpha}_\mu\, {\mathtt{e}^\mu}_\beta={\delta^\alpha}_\beta$. From here, the conserved quantities can be computed. If one replaces 
\begin{align}
   \psi(x)&= \int  \frac{{\rm d}^3 p }{(2\pi)^{3}\sqrt{2 p_0}}\, \sum_{s=1,2} \,\left(a_{\mathbf{p},s}u^{(s)}(\mathbf{p}) e^{-\mathi x^\lambda p_\lambda}+b^\dagger_{\mathbf{p},s}v^{(s)}(\mathbf{p}) e^{\mathi x^\lambda p_\lambda} \right),\nonumber\\
      \psi^\dagger(x)&= \int  \frac{{\rm d}^3 q }{(2\pi)^{3}\sqrt{2 q_0}}\, \sum_{r=1,2} \,\left(b_{\mathbf{q},r}v^{(r)\dagger}(\mathbf{q}) e^{-\mathi x^\lambda q_\lambda}+a^\dagger_{\mathbf{q},r}u^{(r)\dagger}(\mathbf{q}) e^{\mathi x^\lambda q_\lambda} \right)\,,
   \label{eq:diracqdef}
\end{align}
into $ T_{0\mu}$,
one obtains:
\begin{eqnarray}
   && P_\mu=  \int {\rm d}^3 x\,  T_{0\mu}=i \int {\rm d}^3 x\, \bar\psi(x)\, \gamma_0\, \tilde \partial_\mu \psi(x) =i \int {\rm d}^3 x\, \psi^\dagger(x)\,  \tilde \partial_\mu \psi(x) \nonumber\\
   && = -\int  \frac{{\rm d}^3 p \, {\rm d}^3 q \, {\rm d}^3 x}{(2\pi)^{6}\sqrt{4 p_0\, q_0}}\,  p_\mu \, \Omega(p^2) \sum_{r,s=1,2} \,\left(- b_{\mathbf{q},r}\, a_{\mathbf{p},s}\, v^{(r)\dagger}(\mathbf{q}) \, u^{(s)}(\mathbf{p}) \, e^{-\mathi x^\lambda (q_\lambda+p_\lambda)}+ b_{\mathbf{q},r}\, b^\dagger_{\mathbf{p},s}\, v^{(r)\dagger}(\mathbf{q}) \, v^{(s)}(\mathbf{p}) \, e^{-\mathi x^\lambda (q_\lambda-p_\lambda)}\right. \nonumber\\
   &&\left. \hspace{30pt} - a^\dagger_{\mathbf{q},r}\, a_{\mathbf{p},s}\, u^{(r)\dagger}(\mathbf{q}) \, u^{(s)}(\mathbf{p}) \, e^{\mathi x^\lambda (q_\lambda-p_\lambda)}+ a^\dagger_{\mathbf{q},r}\, b^\dagger_{\mathbf{p},s}\, u^{(r)\dagger}(\mathbf{q}) \, v^{(s)}(\mathbf{p}) \, e^{\mathi x^\lambda (q_\lambda+p_\lambda)}\right) \nonumber\\
   &&= - \int  \frac{{\rm d}^3 p }{(2\pi)^{3}2 p_0}\,  p_\mu \,\Omega(p^2) \sum_{r,s=1,2} \,\left(- b_{\mathbf{-p},r}\, a_{\mathbf{p},s}\, v^{(r)\dagger}(\mathbf{-p}) \, u^{(s)}(\mathbf{p}) + b_{\mathbf{p},r}\, b^\dagger_{\mathbf{p},s}\, v^{(r)\dagger}(\mathbf{p}) \, v^{(s)}(\mathbf{p}) - a^\dagger_{\mathbf{p},r}\, a_{\mathbf{p},s}\, u^{(r)\dagger}(\mathbf{p}) \, u^{(s)}(\mathbf{p})\right. \nonumber\\
   &&\left. \hspace{30pt} +  a^\dagger_{\mathbf{-p},r}\, b^\dagger_{\mathbf{p},s}\, u^{(r)\dagger}(\mathbf{-p}) \, v^{(s)}(\mathbf{p}) \right) = \int \frac{{\rm d}^3 p}{(2\pi)^3} \, p_\mu \, \Omega(p^2) \sum_{s=1,2} \left(b^\dagger_{\mathbf{p},s} b_{\mathbf{p},s} +a^\dagger_{\mathbf{p},s} a_{\mathbf{p},s}\right)\, ,
\end{eqnarray}
where we have used the normal ordering for the creation and annihilation operators, the integral definition of $\delta^3(p-q)$, as well as the following orthogonality and normalization rules \cite{Colemanbook}:
\begin{equation}
    u^{(r)\dagger}(-\mathbf{p})\, v^{(s)}(\mathbf{p})=v^{(r)\dagger}(-\mathbf{p})\, u^{(s)}(\mathbf{p})=0, \qquad v^{(r)\dagger}(\mathbf{p})\, v^{(s)}(\mathbf{p})=u^{(r)\dagger}(\mathbf{p})\, u^{(s)}(\mathbf{p})=2p_0 \, \delta_{rs}\, .
\end{equation}
Notice that once we have replace the fields by their modes' decomposition~\eqref{eq:diracqdef}, it can be easily checked that
\begin{eqnarray}
 -\int {\rm d}^3 x\, \eta_{0\mu} \, \bar\psi(x)\left(i\gamma^\alpha \tilde \partial_\alpha -m\right) \psi(x)=0,
\end{eqnarray}
due to the deformed Dirac equations \begin{equation}
(\Omega(p^2) \gamma^\alpha p_\alpha -m)\, u(\mathbf{p})= (\Omega(p^2) \gamma^\alpha p_\alpha +m)\, v(\mathbf{p})=0, \quad \textrm{with} \quad p_\alpha=-i\partial_\alpha.
\end{equation}

\subsection{EM field theory}
The Lagrangian for the EM interaction with a generic metric is given by
\begin{eqnarray}
    \mathcal{L}= -\frac{\sqrt{-\mathtt{g}}}{4} \tilde F_{\rho\sigma}\tilde F^{\rho\sigma}\,.
\end{eqnarray}
Performing small variations of the action with respect to $\mathtt{g}$ one obtains
\begin{eqnarray}
\delta \mathcal{L}=\frac{1}{8} \mathtt{g}_{\mu\nu} \delta \mathtt{g}^{\mu\nu} \tilde F_{\rho\sigma}\tilde F^{\rho\sigma} -\frac{1}{4} \tilde F_{\rho\sigma}\tilde F_ {\mu\nu}\left(\delta \mathtt{g}^{\mu\rho} \mathtt{g}^{\nu\sigma}+ \mathtt{g}^{\mu\rho}\delta\mathtt{g}^{\nu\sigma}\right)=\frac{1}{8} \mathtt{g}_{\mu\nu} \delta \mathtt{g}^{\mu\nu}  \tilde F_{\rho\sigma}\tilde F^{\rho\sigma} -\frac{1}{2} \tilde F_{\mu\sigma}\tilde F_ {\nu\tau}\delta \mathtt{g}^{\mu\nu} \mathtt{g}^{\sigma\tau} \, .
\end{eqnarray}
Consequently, the energy-momentum tensor can be written as
\begin{eqnarray}
    T_{\mu\nu}&=& \left.\frac{2}{\sqrt{-\mathtt{g}}} \, \frac{\delta\mathcal{L}}{\delta \mathtt{g}^{\mu\nu}(x)}\right|_{g_{\mu\nu} \to \eta_{\mu\nu}} =\left. \left[\frac{1}{4} \mathtt{g}_{\mu\nu}  \tilde F_{\rho\sigma}\tilde F^{\rho\sigma} - \tilde F_{\mu\sigma}\tilde F_ {\nu\tau}\mathtt{g}^{\sigma\tau}\right]\right|_{g_{\mu\nu} \to \eta_{\mu\nu}} \nonumber\\
&=&\frac{1}{4} \eta_{\mu\nu}  \tilde F_{\rho\sigma}\tilde F^{\rho\sigma} + \tilde F_{\mu\rho}\tilde F^\rho_{\,\,\,\,\nu} \,.
\end{eqnarray}

 %


\end{document}